\documentclass[aps,twocolumn,showpacs,amsmath,amssymb,superscriptaddress,longbibliography,prb]{revtex4-2}
\usepackage[colorlinks=true,citecolor=blue,linkcolor=blue,hypertexnames=false]{hyperref}
\usepackage{amsmath}
\usepackage{dsfont}
\usepackage{hyperref}
\usepackage{textcomp}
\usepackage{tabularx}
\usepackage{graphicx,tikz}

\newcommand{\cE}{\mathcal{E}}
\newcommand{\bk}{{\bf k}}
\newcommand{\bx}{{\bf x}}
\newcommand{\bq}{{\bf q}}
\newcommand{\ba}{{\bf a}}
\newcommand{\bb}{{\bf b}}

\newcommand{\bP}{{\bf P}}

\newcommand{\bX}{{\bf X}}

\def \be{\begin{equation}}
\def \ee{\end{equation}}
\def \bea{\begin{eqnarray}}
\def \eea{\end{eqnarray}}

\def \be{\begin{equation}}
\def \ee{\end{equation}}
\def \bea{\begin{eqnarray}}
\def \eea{\end{eqnarray}}

\begin{document}

\title{The multi-state geometry of shift current and polarization}

\author{Alexander Avdoshkin}
\thanks{These authors contributed equally.}
\affiliation{Department of Physics, Massachusetts Institute of Technology, Cambridge, MA 02139, USA}
\affiliation{Department of Physics, University of California, Berkeley, CA 94720, USA}

\author{Johannes Mitscherling}
\thanks{These authors contributed equally.} 
\affiliation{Department of Physics, University of California, Berkeley, CA 94720, USA}
\affiliation{Max Planck Institute for the Physics of Complex Systems, N\"othnitzer Str. 38, 01187 Dresden, Germany}

\author{Joel E. Moore}
\affiliation{Department of Physics, University of California, Berkeley, CA 94720, USA}
\affiliation{Materials Sciences Division, Lawrence Berkeley National Laboratory, Berkeley, CA 94720, USA}

\date{\today}

\begin{abstract}
    The quantum metric and Berry curvature capture essential properties of non-trivial Bloch states and underpin many fascinating phenomena. However, it becomes increasingly evident that a more comprehensive understanding of quantum state geometry is necessary to explain properties involving Bloch states of multiple bands, such as optical transitions. To this end, we employ quantum state projectors to develop an explicitly gauge-invariant formalism and demonstrate its power with applications to non-linear optics and the theory of electronic polarization. We provide a simple expression for the shift current that resolves its precise relation to the moments of electronic polarization, clarifies the treatment of band degeneracies, and reveals its decomposition into the sum of the skewness of the occupied states and intrinsically multi-state geometry. The projector approach is applied to calculate non-linear optical properties of transition metal dichalcogenides (TMDs) layers, using previously calculated minimal tight-binding models, and demonstrated analytically on a three-band generalization of the Rice-Mele chain to elucidate the different contributions. We close with comments on  further applications of the projector operator approach to multi-state geometry.
\end{abstract}

\maketitle

\emph{Introduction.---} 
The presence of a crystal lattice with multiple orbitals per unit cell not only leads to (potentially degenerate) band dispersions $E_n(\bk)$ but also, in general, to a non-trivial evolution with crystal momentum $\bk$ of the Bloch states, which captures the spatially periodic part of the lattice eigenstates. The implications of evolving Bloch states have been extensively studied in the emerging field of quantum state geometry, which aims to describe Bloch state properties and their observable consequences using the Berry phase and curvature \cite{Nagaosa2010, Moore2010, Xiao2010}, among other lesser-known quantities. In particular, the quantum metric of Bloch states has attracted much attention in recent years as relevant, for example, for superconductivity \cite{Peotta2015, Hofmann2020, Chen2024} and normal-state transport \cite{Mitscherling2022, Bouzerar2021, Huhtinen2022, Kruchkov2023a} in flat-band materials, the capacitance of insulators \cite{Komissarov2024}, and fractional Chern insulator physics \cite{Roy2014, Mera2021, Ledwith2023}.

The existence and importance of geometrical properties beyond the Berry curvature and quantum metric have also become evident. Besides various generalizations \cite{Niu1985, Mera2022, Chen2022, Kashihara2023, Verma2024}, efforts towards a comprehensive geometric analysis of the Bloch state manifold have revealed new independent geometric quantities involving additional momentum derivatives, multiple momenta, and several band states \cite{Ahn2020, Ahn2022, Bouhon2023, Avdoshkin2022, Avdoshkin2024, Antebi2024}, whose consequences for even uncorrelated electrons in multi-orbital systems are far from completely understood. Linear and non-linear optical responses are natural candidates for the physical manifestation of multi-state quantum geometry due to their momentum-local coupling of the Bloch states of multiple bands \cite{Sipe2000, Fregoso2017, Ahn2022, Zhu2024}. Whereas first-order optical responses only couple to the two-state quantum metric and Berry curvature \cite{Souza2000}, second- and higher-order optical responses involve higher-order tensors \cite{Ahn2020, Ahn2022, Holder2020, Kaplan2023, Wojciech2024} implying geometrical optical sum rules and other results \cite{Souza2000, Kruchkov2023, onishi2024quantum, onishi2024universal, onishi2024fundamental, Verma2024, Wojciech2024}. Multiple-band properties require new concepts of multi-state geometry beyond the broadly studied geometry of single bands or even adiabatic multiband properties like the magnetoelectric effect \cite{qihugheszhang, essinturnermoorevanderbilt, malashevich}. The shift current is an example of the implications of band geometry for technology as it forms one piece of the bulk photovoltaic effect in non-centrosymmetric materials \cite{sturmanbook, Rappe, Cook2017, Alexandradinata2024,Zhu2024}. Hence, a comprehensive understanding of the geometry of optical responses would help guide their identification and their realization in materials.

In this letter, we develop a general approach to computing the geometrical properties of Bloch states, which we apply to shift current and the polarization distribution. Our approach starts from the Bloch Hamiltonian that determines both band dispersions and Bloch states:
\begin{align} \label{eqn:Hdecomp}
    \hat H(\bk) = \sum_{n \,\in\, \text{bands}}\,E_n(\bk)\,\hat P_{n}(\bk) \, ,
\end{align}
where we conveniently express the band Bloch states in a gauge-invariant form via orthogonal projectors $\hat P_n(\bk)$ fulfilling $\hat P_n(\bk)\,\hat P_m(\bk) = \delta_{nm}\hat P_n(\bk)$ \cite{Pozo2020, Graf2021, Mera2021, Mera2022, Avdoshkin2022, Avdoshkin2024, Antebi2024}. The formalism developed in terms of these projection operators enables us to clarify the relation between the shift current and the polarization and unify sum rules in previous special cases~\cite{Fregoso2017, Patankar2018, Zhu2024, agarwal2024shift}. It also solves challenges in the treatment of degenerate bands \cite{Sipe2000}, which we combine with a complete geometric characterization of the polarization of a generic band insulator. Generalizing previous single-band results \cite{Souza2000, Avdoshkin2022}, we identify the geometry associated with the Slater determinant ground state of uncorrelated electron systems, presenting Pl\"ucker embedding (essentially a formalization of Slater determinants) as a convenient description for degenerate subspaces \cite{Avdoshkin2024}. An advantage of the formulas based on projectors is their explicit gauge invariance, and we apply full and simplified versions of the resulting expressions for optical properties to three-band models of transition metal dichalcogenides (TMDs) \cite{Liu2013} to demonstrate their practical utility. In the companion work \cite{Mitscherling2024}, we complement the physical insights with considerably more technical details on the projector formalism and the geometrical invariants focusing on broadly relevant algebraic relationships and detailed example calculations. 

\emph{Geometry of shift current.---} The DC components of the second-order optical response under uniform illumination 
\begin{align}
    \label{eqn:sigma_def}
    j^\alpha(0) = \sum_{\beta,\gamma}\,\Big(\sigma^{\alpha;\beta\gamma}_\text{inj}(\omega)+\sigma^{\alpha;\beta\gamma}_\text{shift}(\omega)\Big)\, \cE^\beta(\omega) \,\cE^\gamma(-\omega) \, 
\end{align}
decomposes into the injection and shift current with Fourier components of the electric fields $\cE^\beta(\omega)$ \cite{Sipe2000, Ahn2020, Holder2020, Ahn2022, Kaplan2023}. The injection current involves the two-state (second-order) quantum geometric tensor $Q^{mn}_{\beta\gamma}$ \cite{Ahn2020}, whereas the shift current is determined by the two-state third-order quantum geometric connection $C^{mn}_{\alpha;\beta\gamma}$ \cite{Ahn2020, Ahn2022}, whose symmetric and antisymmetric components determine the linear and circular shift current \cite{SupplMat} via
\begin{alignat}{3}
    \label{eqn:sigma_linear}
    &\sigma^{\alpha;(\beta\gamma)}_\text{shift}(\omega)&&
    \!=\!-\frac{2\pi e^3}{\hbar^2}\!\!\!\sum_{\underset{n<m}{n,m}}\!\int_\bk\!\!\delta(\omega-\epsilon_{mn})f_{nm}\,\text{Im}\,C^{[mn]}_{\alpha;(\beta\gamma)} \,,\\
    \label{eqn:sigma_circular}
    & \sigma^{\alpha;[\beta\gamma]}_\text{shift}(\omega)&&
    \!=\!-\frac{2i\pi e^3}{\hbar^2}\!\!\!\sum_{\underset{n<m}{n,m}}\!\int_\bk\!\!\delta(\omega-\epsilon_{mn})f_{nm}\text{Re}\,C^{(mn)}_{\alpha;[\beta\gamma]}  
    \,,
\end{alignat}
where we indicate (anti-)symmetrization in the indices as $(\beta\gamma)$ and $[\beta\gamma]$. We assumed $\omega > 0$ and bands sorted by increasing energy. $\epsilon_{nm}\equiv E_n-E_m$ and $f_{nm}=f_n-f_m$ denote the band gaps and Fermi function differences. We omit the momentum dependence for brevity. The momentum integral reads $\int_\bk\equiv \int_\text{BZ}d^d\bk/(2\pi)^d$ in $d$ dimensions and the electron's charge is $-e$ with $e>0$.

\emph{Multi-state geometry.---} 
The second- and third-order geometric tensors $Q^{mn}_{\beta\gamma}$ and $C^{mn}_{\alpha;\beta\gamma}$ in projector form read
\begin{align}
    \label{eqn:Qmn_bc}
    &Q^{mn}_{\beta\gamma}\,\equiv \text{tr}\Big[\hat P^{}_{n}\,(\partial^{}_\beta\hat P^{}_{m})\,(\partial^{}_\gamma\hat P^{}_{n})\Big] \, , \\
    \label{eqn:Cmn_abc}
    &C^{mn}_{\alpha;\beta\gamma}
    \!\equiv\text{tr}\Big[\hat P^{}_{n}(\partial^{}_\beta\hat P^{}_{m})\big[(\partial^{}_\alpha\partial^{}_\gamma\hat P^{}_{n})\!+\!(\partial^{}_\alpha\hat P^{}_{m})(\partial^{}_\gamma\hat P^{}_{n})
    \big]\!\Big] ,
\end{align}
with momentum derivatives $\partial_\alpha\equiv \partial_{k_\alpha}$. The diagonal component of $Q^{nn}_{\beta\gamma} = g^n_{\beta\gamma}-\frac{i}{2}\Omega^n_{\beta\gamma}$ yields the quantum metric and Berry curvature of band $n$ as real and imaginary part. The single-band component of $C^{nn}_{\alpha;\beta\gamma} = \text{tr}[\hat P_n(\partial_\beta \hat P_n)(\partial_\alpha\partial_\gamma\hat P_n)]$ is a novel independent extrinsic geometric quantity of band $n$ \cite{Avdoshkin2022}. 
The two terms in Eq.~\eqref{eqn:Cmn_abc} are different from the separation used to clarify the real-space and momentum-space parts of the shift vector \cite{kaplan2022} as both terms are separately gauge-invariant.

A key reason to write the geometric objects in the form of  Eqs.~\eqref{eqn:Qmn_bc} and \eqref{eqn:Cmn_abc} is that they generalize naturally from two states to two subspaces: replacing single-band projectors by subspace projectors, e.g., double-degenerate bands $\hat P_{n} = \hat P_{(n1)} + \hat P_{(n2)}$ or $\hat P_\text{occ}=\sum_{n\in \text{occ}} \hat P_n$ for the subspace of occupied bands, the $\bigotimes_{n} \!U(N_n)$-gauge invariance under unitary transformations of the $N_n$-degenerate subspaces is evident in this formalism by gauge-invariance of each projector. The gauge invariance shortens analytic expressions, removing unphysical redundancies, and allows for a straightforward numerical evaluation \cite{Mitscherling2024} as the projectors can be chosen to be smooth over the whole Brillouin zone away from band crossings, unlike the Bloch states \cite{Mera2022}.

In previous derivations of the injection and shift currents \cite{Sipe2000, Ahn2020, Ahn2022}, the treatment of degenerate bands remained challenging. In Ref.~\onlinecite{Sipe2000}, degeneracies of the symmetric contribution $\sigma^{\alpha;(\beta\gamma)}_\text{shift}$ were treated in the basis where degenerate bands are split according to $\sum_\alpha\cE^\alpha\langle u_{(ns)} |\partial_\alpha u_{(nl)}\rangle = 0$ for $s\neq l$, where $s, l$ label different Bloch states $|u_{(ns)}\rangle$ within the degenerate band $n$. More recently, gauge invariance was assured by summing over the degenerate subspace explicitly \cite{Ahn2020, Ahn2022}. The projector formulation in Eq.~\eqref{eqn:Qmn_bc} and \eqref{eqn:Cmn_abc} suggests a treatment of degenerate and non-degenerate bands on an equal footing by only adjusting the corresponding band projectors. Indeed, we show by explicit calculation presented in the companion paper \cite{Mitscherling2024} that the derivation of Eq.~\eqref{eqn:sigma_def} only requires the Hamiltonian decomposition given in Eq.~\eqref{eqn:Hdecomp}, where the projectors $\hat P_n = \sum^{N_n}_{s=1} \hat P_{(ns)}$ may project onto $N_n$-degenerate band subspaces. We note that finite intra- and interband relaxation rates \cite{Holder2020} enable further contributions to the shift current \cite{Kaplan2023, Mitscherling2024} and allow the inclusion of effectively degenerate bands \cite{Mera2022}. 

While the separation of the two-state quantum metric and Berry curvature as symmetric and antisymmetric parts of $Q^{mn}_{\beta\gamma}$ is well established, the quantum geometric connection yields a more complex structure, revealed by symmetrization of the band and external indices \cite{SupplMat}, 
\begin{align}
    \label{eqn:Cmn_decomp}
    C^{mn}_{\alpha;\beta\gamma}=&\frac{1}{2}\partial^{}_\alpha\,Q^{mn}_{\beta\gamma}+\text{Re}\,C^{(mn)}_{\alpha;[\beta\gamma]}+i\,\text{Im}\,C^{[mn]}_{\alpha;(\beta\gamma)} 
\end{align}
for $n\neq m$. Note the connection between the symmetry in the band indices and being purely real and imaginary. Interestingly, given as the first term, the quantum metric and Berry curvature dipole \cite{Sodemann2015, Wang2023} are part of the tensor. The linear and circular shift currents provide the remaining information about $C^{mn}_{\alpha;\beta\gamma}$. 

The shift current can be related to a generalized Wilson loop associated with interband transitions at separated momenta \cite{Shi2021, Wang2022, Zhu2024}. This description is expressible in the projector approach by defining the interband Wilson loop as $W^{mn}_{\alpha\beta}(\bk,\bq) = \text{tr}\big[\hat e^{nm}_\beta(\bk)\,\hat e^{mn}_\alpha(\bk+\bq)\big]$ with the gauge-invariant interband dipole moments in projector form $\hat e^{mn}_\alpha = i\hat P_m\big(\partial_\alpha \hat P_m\big)\hat P_m$ \cite{Ahn2022, Mitscherling2024}. The Wilson loop representation of the shift vector \cite{Wang2022} leads to the projector form $R^{mn}_{\alpha\beta}(\bk) = i\,C^{mn}_{\alpha;\beta\beta}(\bk)/Q^{mn}_{\beta\beta}(\bk)$, which reduces to the conventional definition \cite{Sipe2000, Ahn2022} for non-degenerate bands \cite{Mitscherling2024}.

\emph{Geometric sum rules.---} Eqs.~\eqref{eqn:sigma_linear} and \eqref{eqn:sigma_circular} imply that the linear and the circular shift current are, after frequency integration, entirely independent of band energies for an insulator and, thus, purely geometric. Therefore, they resemble the well-known Souza-Wilkens-Martin sum rule, which relates the optical conductivity to the integrated quantum metric of the occupied states interpreted as the ground state spread in real space \cite{Souza2000}.

One might wonder whether it is possible to generally reduce $C^{mn}_{\alpha;\beta\gamma}$ to the projector $\hat P_\text{occ}$ on all occupied bands upon band summation. Surprisingly, this is not possible for systems with more than two bands. As an illustration, we consider a single (potentially degenerate) filled band $\hat P_\text{occ} = \hat P_0$. The summation of the unoccupied subspace leads to 
\begin{align}
    \label{eqn:sum_Cmn}
    \sum_{m\in \text{unocc}}\!\!\!\! C^{m0}_{\alpha;\beta\gamma} =&  - \!\text{tr}\big[\hat P_\text{occ}\,\big(\partial_\beta \hat P_\text{occ}\big)\big(\partial_\alpha\partial_\gamma \hat P_\text{occ}\big)\big] \nonumber \\ &\hspace{-14mm}+ \!\text{tr}\Big[\hat P_\text{occ} \Big(\!\sum_{m \in \text{unocc}} \!(\partial_\beta \hat P_m) (\partial_\alpha \hat P_m) \!\Big)(\partial_\gamma\hat P_\text{occ})\Big] \, ,
\end{align}
where the generically nonzero second term involves the sum over all unoccupied bands. Further terms arise when the occupied subspace involves multiple bands at different energies \cite{SupplMat}. Thus, the shift current sum rule does not reduce to a ground state property of the occupied subspace, unlike the Souza-Wilkens-Martin sum rule \cite{Souza2000}.

We can relate part of the deviation from a ground state property to the torsion tensor, a purely two-state geometric quantity. Building on the previous definition \cite{Ahn2022, Wojciech2024}, we express torsion in terms of the two subspace projectors $\hat P_n$ and $\hat P_m$, 
\begin{align}
    \label{eqn:torsion}
    T^{mn}_{\beta;\alpha\gamma} = \text{tr}\big[\hat P_n (\partial_\beta \hat P_m)(\partial_\alpha\hat P_m)(\partial_\gamma\hat P_n)\big]-(\alpha\leftrightarrow\gamma) \, .
\end{align}
It is symmetric in its last two indices and vanishes for $n=m$ \cite{SupplMat}. The symmetrized sum rule of the circular shift current in Eq.~\eqref{eqn:sigma_circular} relates to the real part of the momentum-integrated and band-traced torsion via the second term in Eq.~\eqref{eqn:sum_Cmn} \cite{Wojciech2024, SupplMat} and is quantized in some real-valued 3- and 4-band models \cite{Wojciech2024, jankowski2024non}. We expect further purely multi-state geometric quantities to be essential for higher-order optical responses, which remain to be identified and characterized.

\begin{figure*}[t!]
    \centering
    \includegraphics[width=0.23\linewidth]{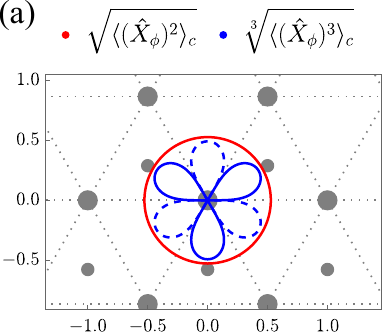}
    \includegraphics[width=0.23\linewidth]{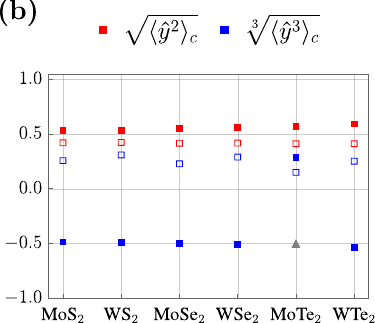}
    \includegraphics[width=0.23\linewidth]{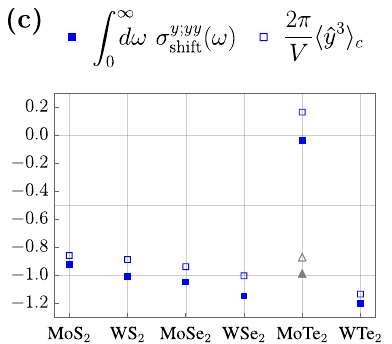}
    \includegraphics[width=0.23\linewidth]{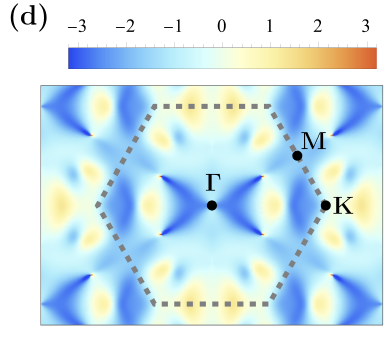}
    \caption{(a) The roots of the variance (red) and skewness (blue) of the filled lowest band in units of the lattice constant for the transition metal dichalcogenide (TMD) $\text{MoS}_2$ using TNN-GGA parameters in relation to the triangular lattice of the tight-binding model (big dots). The dashed blue line indicates negative skewness.
    (b) The roots of the variance (red) and skewness (blue) for different TMDs comparing NN- and TNN-GGA parameters (open and filled boxes). Whereas the skewness for $\text{MoTe}_2$ strongly deviates using GGA parameters, a compatible value is found for LDA parameters (gray triangle).
    (c) The skewness of the filled band (empty boxes) yields the dominant contribution to the linear shift current sum rule (filled boxes) given in units $e^3/\hbar^2$. A clear trend for the different TMDs using TNN-GGA parameters is evident, where the TNN-LDA parameters for $\text{MoTe}_2$ (gray triangles) lead to better consistency.
    (d) The momentum-resolved contribution to the linear shift current sum rule for $\text{MoS}_2$ using TNN-GGA parameters, where we indicate the Brillouin zone (dashed lines) and high symmetry points.
    }
    \label{fig:TMDresults}
\end{figure*}

\emph{Polarization distribution.---} We take a closer look at the first term in Eq.~\eqref{eqn:sum_Cmn} and establish its relationship to the modern theory of polarization \cite{Kingsmith1993, Resta2007}. The primary tool in this construction is the reduction of the geometry of the occupied bands captured by $\hat P_\text{occ}$ to the geometry of a single state $|\Psi\rangle$ via the Pl\"ucker embedding, whose details we carefully discuss in the companion paper \cite{Mitscherling2024}. For this, we consider the polarization operator of all electrons $\hat \bP = -e \,\hat \bX$ involving the many-body position operator $\hat{\bX} = \sum_{i\in\text{sites}} \hat{\bx}_i$. The generating function of its moments reads
\begin{align}
    C(\bq) = \langle \Psi |e^{i \bq\cdot\hat{\bX}}| \Psi \rangle 
\end{align}
concerning the ground state wave function $| \Psi \rangle$. Assuming that $|\Psi\rangle$ is a Slater determinant constructed out of occupied bands, we obtain a closed formula for all cumulants of the distribution \cite{Mitscherling2024}
\begin{align} \label{eq:gen_func}
    \frac{\log C(\bq)}{V} = \sum_\alpha q^\alpha\mathcal{A}_\alpha+ \sum_\alpha q^\alpha \!\!\int_\bk \int_0^1\!\!\!\!dt\,\, \mathcal{A}_\alpha^{\bk}(\bk+\bq t) \, .
\end{align}
where the constant $\mathcal{A}_\alpha = \sum_{n\in\text{occ}}\int_\bk \mathcal{A}^n_\alpha(\bk)$ is fixed by the known relation between the mean polarization and the Berry connection \cite{Resta2007}. All higher moments are given by $\mathcal{A}_{\alpha}^{\bk}(\bk+\bq) = \text{tr}\Big[\hat P_{\bk}\big(\hat P_{\bk} \hat P_{\bk+\bq}\hat P_{\bk}\big)^{-1} \hat P_{\bk}\big(\partial_{\alpha}\hat P_{\bk+\bq}\big) \hat P_{\bk+\bq} \Big]$ involving the projector $\hat P_\bk \equiv \hat P_\text{occ}(\bk)$ on all occupied bands for different $\bk$. This result generalizes the single-band formula given in Ref.~\onlinecite{Avdoshkin2022} to an arbitrary number of occupied bands.

Expanding Eq.~\eqref{eq:gen_func} up to second and third order in $\bq$, we obtain the cumulants
\begin{align}
    \langle \hat X_\alpha \hat X_\beta \rangle_c &=  V \!\!\!\int_\bk \!\!\text{Re}\,\text{tr}\Big[\hat P_\text{occ} \big(\partial_\alpha \hat P_\text{occ}\big)\!\big(\partial_\beta \hat P_\text{occ}\big)\!\Big] \, ,\label{eqn:variance} \\ 
    \langle \hat X_\alpha \hat X_\beta \hat X_\gamma \rangle_c  &= V \!\!\!\int_\bk \!\!\text{Im}\,\text{tr}\Big[\hat P_\text{occ} \big(\partial_\alpha \hat P_\text{occ}\big)\! \big(\partial_\beta \partial_\gamma \hat P_\text{occ}\big)\!\Big] \,,\label{eqn:skewness}
\end{align}
where $V$ is the volume of the unit cell and $\langle \hat X_\alpha^2\rangle_c = \langle \big(\hat X_\alpha-\langle \hat X_\alpha\rangle\big)^2 \rangle$ and $\langle \hat X_\alpha^3\rangle_c = \langle \big(\hat X_\alpha-\langle \hat X_\alpha\rangle\big)^3 \rangle$ when focused on a single spatial direction $\alpha$.

By comparing Eq.~\eqref{eqn:sum_Cmn} with Eq.~\eqref{eqn:variance}, we obtain that the integrated linear shift current is proportional to the skewness of the occupied states for two-band systems \cite{Patankar2018}, but multi-state geometry (i.e., the second term in Eq.~\eqref{eqn:sum_Cmn}) leads to deviations in systems with more than two bands. An approximate ground-state sum rule is obtained by dropping the multi-band contribution,
\begin{align}
    \label{eqn:sumrule}
    \int_0^\infty \!\!\!\!\!d\omega\,\,\sigma^{\alpha;(\beta\gamma)}_\text{shift}(\omega) \approx \,&\frac{2\pi e^3}{\hbar^2} \frac{1}{V}\langle \hat X_\alpha \, \hat X_\beta \, \hat X_\gamma\rangle_c \, .
\end{align}
The simplified sum rule serves as an estimate for the strengths of a material's linear shift current response in terms of a straightforward ground state property, the skewness. In the next section, we quantify the accuracy of this simplification for a practical example.

\emph{Illustration for the three-band model of TMDs. ---} 
We consider a single-layer of transition metal dichalcogenides (TMDs) $\text{MX}_2$ with $\text{M} = \text{Mo},\text{W}$ and $\text{X}=\text{S},\text{Se},\text{Te}$ (see Fig.~\ref{fig:TMDresults} (a)). The M atoms (big dots) form a triangular lattice used within the three-band tight-binding Hamiltonian proposed by Liu {\it et al.} \cite{Liu2013}. We use their model parameters for nearest-neighbor (NN) and next-nearest-neighbor hoppings (TNN) calculated by ab initio density functional theory (DFT) using generalized-gradient approximation (GGA) and local-density approximation (LDA). The resulting band structure shows a lower, isolated, non-degenerate band, which we assume to be filled. The NN models capture the DFT band structure around $\text{K}=(4\pi/3,0)$, whereas the TNN models provide an overall good agreement for all momenta \cite{Liu2013}. We provide the details of the numerical evaluation in supplemental material \cite{SupplMat}. We work in the widely used minimal tight-binding approximation, in which the off-diagonal matrix elements of the position operator are neglected; improved accuracy can be obtained either by increasing the number of bands or including off-diagonal elements \cite{dejuan2022,Zhu2024}.

We calculate $\langle (\hat X_\phi)^2\rangle_c$ and $\langle (\hat X_\phi)^3\rangle_c$ via Eqs.~\eqref{eqn:variance} and \eqref{eqn:skewness} and show their second and third root in spatial direction $\hat X_\phi \equiv \sin(2\pi \phi)\,\hat x + \cos(2\pi\phi)\,\hat y$ for $\text{MoS}_2$ centered at an M atom, which enables their comparison with the underlying lattice, see Fig.~\ref{fig:TMDresults} (a), with lattice constant set as reference scale. The variance (red) is directionally independent, whereas the $C_3$ rotational symmetry becomes evident in the skewness (blue), which captures the positive tilt of the occupied states towards the nearest X atoms. We compare their amplitudes for six TMDs, sorted by increasing lattice constant, in Fig.~\ref{fig:TMDresults} (b). We see that the variance and the skewness are smaller in all NN models, which suggests a stronger M-X hybridization in the TNN models. Most importantly, the skewness obtained from the NN models has an opposite sign compared to those of the TNN models, which suggests that the longer-range hoppings over the X atoms (small dots in Fig.~\ref{fig:TMDresults} (a)) are not only required for the band dispersion but also necessary to capture essential quantum state properties due to the X atoms. As the skewness and shift current are closely related, we conclude that the NN models of TMDs are insufficient for an adequate description of the shift current and its sum rule. 

We note that the result of $\text{MoTe}_2$ strongly deviates from the other TMDs. Since the skewness in the LDA model (gray triangle) is in line with the other TMDs, perhaps the determination of the tight-binding parameters failed for GGA, setting $\text{MoTe}_2$ at a potentially interesting parameter range. This discrepancy may be related to the band crossing on the M-$\Gamma$ high-symmetry line seen in DFT between the filled band and a lower band not included in the three-band description \cite{Liu2013}. 

To verify the accuracy of the simplified sum rule in Eq.~\eqref{eqn:sumrule}, we show $\int_0^\infty \!d\omega \,\sigma^{y;yy}_\text{shift}(\omega)$ (filled boxes) in comparison to the skewness in units $e^3/\hbar^2$ in Fig.~\ref{fig:TMDresults} (c) and find good agreement. The multi-state geometry leads only to minor contributions, such that the skewness of the occupied lowest band contributes approximately $90\%$ to the linear shift current sum rule for all TMDs. Particularly good agreement is found for $\text{MoS}_2$ and $\text{WTe}_2$. As before, only LDA model parameters for $\text{MoTe}_2$ follow the overall trend, whereas GGA parameters suggest a vanishing sum rule. However, we note that the multi-state geometry significantly alters the momentum distribution of the sum rule integrand, as shown for $\text{MoS}_2$ using TNN-GGA model parameters in Fig.~\ref{fig:TMDresults}(d). The contributions at the Brillouin zone boundary (dashed lines) are mainly due to the momentum-resolved skewness. The structure around $\Gamma$ is related to the multi-state geometry \cite{SupplMat}. The physical divergences on the high-symmetry line $\Gamma$-M correspond to the band crossings of the two conduction bands \cite{SupplMat}, which has no major impact on the integrated linear shift current.

\emph{Stub Rice-Mele model. ---} To shed light on the physical origin of the multi-band contributions in Eq.~\eqref{eqn:sum_Cmn} and the validity of the sum rule \eqref{eqn:sumrule} beyond the two-band limit, we combine the Rice-Mele model \cite{Rice1982} with equally distant sites A and B with locally attached stubs C \cite{Neves2024}. In the Bloch basis $(c^\dagger_{k,A}, c^\dagger_{k,B}, c^\dagger_{k,C})$ the Bloch Hamiltonian reads
\begin{align}
    \hat H(k) = \begin{pmatrix} m & 2\cos\frac{k}{2}+2i\delta\sin \frac{k}{2} & \Delta \\ 2\cos \frac{k}{2}-2i\delta\sin \frac{k}{2} & - m & 0 \\ \Delta &  0 & 0 \end{pmatrix} \, .
\end{align}
Both the energy splitting $m$ between sites A and B and the hybridization $\Delta$ between site A and the stub C are mechanisms for a finite skewness as long as the A-B chain dimerizes $(\delta\neq 0)$~\cite{SupplMat}. However, only the latter three-site mechanism leads to multi-band contributions in Eq.~\eqref{eqn:sum_Cmn}, suggesting that the accuracy of the sum rule for monolayer TMDs in Fig.~\ref{fig:TMDresults}~(c) arises from a dominantly two-site mechanism. The toy model indicates that multi-band contributions can enhance, suppress, and even invert non-linear optical responses. Indeed, an enhanced shift current from virtual multiband transitions has recently been suggested for alternating angle twisted multilayer graphene \cite{Chen2024shift}.

\emph{Conclusions and outlook. ---} We have developed a comprehensive geometric approach to observables involving multiple band states, such as optical responses. Applying this theory to the shift current resolves the treatment of band degeneracies and establishes the precise connection between the linear shift current sum rule and the skewness of the ground state. The gauge-invariant projector formalism and techniques for its implementation, whose technical aspects are introduced and analyzed in detail in the companion paper \cite{Mitscherling2024}, will significantly simplify the analytic treatment of higher-order DC and optical response functions in electrical fields with charge and other current vertices, such as thermal. Responses involving the magnetic field and magnetization require a more detailed analysis due to intrinsically broken translational invariance. Another possible extension is to bands modified through light-matter coupling \cite{Topp2019, Topp2021, Tai2023, Walicki2024}. We emphasize that the multi-state projector formalism does not rely on an {\it a priori} projection onto the low-energy sector, which simplifies a systematic identification of all leading-order contributions from real and virtual interband processes \cite{Antebi2024} and generalizes straightforwardly to some symmetry-broken phases \cite{Porlles2023}. Combining the presented approach with current attempts to avoid explicitly fixing quantum states via expansions in generators of SU(N) \cite{Pozo2020, Graf2021, Mitscherling2024} allows for closed analytic forms of the geometric observables for more than two bands, which are needed to show pure multi-state geometric effects.

The geometric analysis of low-energy tight-binding models offers a more refined characterization of the described quantum states. We revealed inconsistencies in ab initio models for TMDs, whose detailed analysis might lead to more refined low-energy models. Our analysis suggests that higher-order quantum geometric tensors are necessary, complementing current characterizations based on the quantum metric \cite{Hirschmann2024} and Berry curvature \cite{Jorissen2024}. In particular, we propose to explore the validity of effective models via the amplitude and angle dependence of geometric sum rules, where the Souza-Wilkens-Martin sum rule yields information on the ground state variance \cite{Souza2000} and the linear shift current sum rule yields information about the ground state skewness. 

Studying the effect of interactions will require understanding global multi-state (multi-momentum) geometric objects, see e.g. the perturbative treatment of interaction in \cite{Antebi2024}. Two promising avenues for investigation include the impact of interactions on quantization \cite{Avdoshkin2020} and the consequences of electron-phonon coupling \cite{PhysRevLett.130.226302,chen2024gauge}. Such analyses have the potential to provide new insights into the role of vertex corrections in optical responses, which pose significant challenges for numerical methods in strongly correlated systems. 

The current approach primarily aims to capture the (higher-order) local invariant objects. Note that some global geometric quantities, e.g., topological invariants, remain challenging. For example, the Zak phase of a band allows a simple representation via the Berry connection, $\phi = \int A$ and can be computed via the product of projectors along loops, but expressing it by local invariant tensors is non-trivial (Appendix G of \cite{Avdoshkin2020}). In contrast, the multi-band magneto-electric coupling has a Chern-Simons part for which a projector expression is not known and no local, invariant representation has been found \cite{Taherinejad2015} without an extra dimension.

In conclusion, a comprehensive understanding of how quantum geometry determines observables will enable a more refined state tomography of quantum materials using a minimal set of observables, of which optical responses are a promising class. The formalism we presented in this work provides a systematic construction of the relevant geometric quantities with a given number of external indices, whose relevance has become evident from the growing number of higher-order geometric quantities \cite{Mitscherling2020, Kozii2021, Sodemann2015, Ahn2022, Wang2023, Fang2024, Kaplan2023}. Having identified a comprehensive set of quantum geometric tensors of different ranks, we can seek to determine the most promising observables via a careful analysis of their symmetries. 

\emph{Acknowledgements.---} We thank 
Aris~Alexandradinata,
Alex Tsigler,
Ohad~Antebi,
Dan~S.~Borgnia,
Iliya~Esin,
Moritz~M.~Hirschmann, 
Tobias~Holder,
Fernando de Juan,
Wojciech~Jankowski,
Libor~\v Smejkal, and
Tha\'is~V.~Trevisan
for stimulating discussions. A.~A. was supported by a Kavli ENSI fellowship at UC Berkeley and the National Science Foundation (NSF) Convergence Accelerator Award No. 2235945 at MIT. J.~M. was supported by the German National Academy of Sciences Leopoldina through Grant No. LPDS 2022-06 and, in part, by the Deutsche Forschungsgemeinschaft under Grant cluster of excellence ct.qmat (EXC 2147, Project No. 390858490). J.~E.~M. was supported by the Quantum Materials program under the Director, Office of Science, Office of Basic Energy Sciences, Materials Sciences and Engineering Division, of the U.S. Department of Energy, Contract No. DE-AC02-05CH11231.at Lawrence Berkeley National Laboratory, and a Simons Investigatorship.

\bibliography{main_arxiv_v2}

\newpage

\clearpage

\begin{widetext}

\begin{center}
\textbf{\large Supplemental Materials for \\[1mm] ``The multi-state geometry of shift current and polarization''} 
\end{center}
\setcounter{equation}{0}
\setcounter{figure}{0}
\setcounter{table}{0}

\makeatletter
\renewcommand{\theequation}{S\arabic{equation}}
\renewcommand{\thefigure}{S\arabic{figure}}
\renewcommand{\thetable}{S\Roman{table}}

\vspace{1cm}
In this Supplemental Material (SM), we provide the details on (I) the symmetry decomposition of the shift current and (II) the separation of the third-order geometric tensor $C^{mn}_{\alpha;\beta\gamma}$. We present (III) the full expression for the summation over band indices of $Q^{mn}_{\beta\gamma}$ and $C^{mn}_{\alpha;\beta\gamma}$ and (IV) further details on the torsion tensor. We summarize (V) all details of the application to transition metal dichalcogenides (TMDs) and close by discussing (VI) the Stub Rice-Mele model as a minimal three-band model for skewness and the corresponding interband corrections.

\section{Symmetry decomposition of the shift current}

We consider the shift current expressed in terms of the multiband geometric quantity as defined in Eq.~\eqref{eqn:Cmn_abc},
\begin{align}
    \label{eqn:formula_shift}
    \sigma^{\alpha;\beta\gamma}_\text{shift}(\omega)=\frac{i\pi e^3}{\hbar^2}\sum_{m,n}\int_\bk\!\delta(\omega-\epsilon_{mn})\,f_{nm}\,(C^{mn}_{a;\gamma\beta}-C^{nm}_{\alpha;\beta\gamma}) \, .
\end{align}
Here, we use the convention of electrical charge $-e$ with $e>0$. The Fermi distribution function reads $f_n\equiv \big[1+e^{(E_n-\mu)/T}\big]^{-1}$ with temperature $T$ setting $k_B=1$, chemical potential $\mu$, and band dispersion $E_n$. We use the short notations $f_{nm}=f_n-f_m$ and $\epsilon_{mn}=E_m-E_n$. In the following, we decompose the shift current into its symmetry components.

As defined in Eq.~\eqref{eqn:sigma_def}, the index $\alpha$ corresponds to the spatial direction of the current, whereas the indices $\beta$ and $\gamma$ label the particular directions of the electric field. Symmetrizing these latter indices allows us to distinguish the contributions related to linear and circular fields. Thus, we define
\begin{align}
    \sigma^{\alpha;\beta\gamma}_\text{shift}(\omega)=\sigma^{\alpha;(\beta\gamma)}_\text{shift}(
    \omega)+\sigma^{\alpha;[\beta\gamma]}_\text{shift}(\omega)
\end{align}
where $(\beta\gamma)$ indicates the symmetrization and $[\beta\gamma]$ indicates antisymmetrization. We include a normalization factor of $1/2$. We note that the symmetrization in both the external indices $\beta\leftrightarrow\gamma$ and the band indices $n\leftrightarrow m$ of the geometric part within Eq.~\eqref{eqn:formula_shift} leads to
\begin{align}
    C^{mn}_{a;\gamma\beta}-C^{nm}_{\alpha;\beta\gamma}=2\Big(C^{[mn]}_{\alpha;(\beta\gamma)}-C^{(mn)}_{\alpha;[\beta\gamma]}\,\Big) \, .
\end{align}
Thus, the symmetry decomposition of the external indices implies a symmetry decomposition in the band indices. Defining
\begin{align}
    W^{mn}(\omega)=\delta(\omega-\epsilon_{mn})\,f_{nm}
\end{align}
and symmetrizing with respect to $n\leftrightarrow m$ by using $f_{nm}=-f_{mn}$ we obtain
\begin{align}
    &W^{(mn)}(\omega)=\frac{1}{2}\big(\delta(\omega-\epsilon_{mn})-\delta(\omega+\epsilon_{mn})\big)\,f_{nm} \, ,\\ 
    &W^{[mn]}(\omega)\,=\frac{1}{2}\big(\delta(\omega-\epsilon_{mn})+\delta(\omega+\epsilon_{mn})\big)\,f_{nm} \, .
\end{align}
This decomposition implies the symmetry for the external frequency, that is, $W^{(mn)}(-\omega) = -W^{(mn)}(\omega)$ and $W^{[mn]}(-\omega) = W^{[mn]}(\omega)$. The shift current reads
\begin{align}
    \sigma^{\alpha;\beta\gamma}_\text{shift}(\omega)&=\frac{2\pi i\, e^3}{\hbar^2}\sum_{m,n}\int_\bk\Big[W^{(mn)}(\omega)+W^{[mn]}(\omega)\Big]\,\Big[C^{[mn]}_{\alpha;(\beta\gamma)}-C^{(mn)}_{\alpha;[\beta\gamma]}\Big]\\
    &=\frac{2\pi i\, e^3}{\hbar^2}\sum_{m,n}\int_\bk\Big[W^{[mn]}(\omega)\,\,C^{[mn]}_{\alpha;(\beta\gamma)}-W^{(mn)}(\omega)\,\,C^{(mn)}_{\alpha;[\beta\gamma]}\Big] \, .
\end{align}
For a convenient evaluation, we label the bands in order of increasing energy and restrict the band summation accordingly. We note that $n\neq m$ due to $f_{nn} = 0$. Furthermore, we use the relation between the symmetry of the band indices and the real and imaginary parts 
and obtain
\begin{align}
    \sigma^{\alpha;\beta\gamma}_\text{shift}(\omega)&=\frac{4\pi i\, e^3}{\hbar^2}\sum_{\underset{n<m}{n,m}}\int_\bk\Big[W^{[mn]}(\omega)\,\,C^{[mn]}_{\alpha;(\beta\gamma)}-W^{(mn)}(\omega)\,\,C^{(mn)}_{\alpha;[\beta\gamma]}\Big]\\
    &=-\frac{4\pi \, e^3}{\hbar^2}\sum_{\underset{n<m}{n,m}}\int_\bk\Big[W^{[mn]}(\omega)\,\,\text{Im}\,C^{mn}_{\alpha;(\beta\gamma)}+i\,W^{(mn)}(\omega)\,\,\text{Re}\,C^{mn}_{\alpha;[\beta\gamma]}\Big] \, .
\end{align}
In combination with the symmetry with respect to $\omega\leftrightarrow -\omega$ of the two contributions $\sigma^{a(\beta\gamma)}_\text{shift}(\omega)=\sigma^{a(\beta\gamma)}_\text{shift}(-\omega)$ and $\sigma^{a[\beta\gamma]}_\text{shift}(\omega)=-\sigma^{a[\beta\gamma]}_\text{shift}(-\omega)$ it is sufficient to restrict ourselves to positive frequency $\omega > 0$ only. By the convention of our band labeling, we have 
$\epsilon_{mn}>0$ so that the two contributions to the shift current take the convenient and compact form
\begin{alignat}{3}
    &\sigma^{\alpha;(\beta\gamma)}_\text{shift}(\omega)&&
    =\,-\frac{2\pi e^3}{\hbar^2}\sum_{\underset{n<m}{n,m}}\int_\bk\delta(\omega-\epsilon_{mn})\,f_{nm}\,\text{Im}\,C^{mn}_{\alpha;(\beta\gamma)} \, , \\
    & \sigma^{\alpha;[\beta\gamma]}_\text{shift}(\omega)&&
    =-\frac{2i\pi e^3}{\hbar^2}\sum_{\underset{n<m}{n,m}}\int_\bk\delta(\omega-\epsilon_{mn})\,f_{nm}\,\text{Re}\,C^{mn}_{\alpha;[\beta\gamma]} \, .
\end{alignat}
In the main text, we provide these formulas in Eqs.~\eqref{eqn:sigma_linear} and \eqref{eqn:sigma_circular}. The formulas explicitly show that the symmetric contribution is real, whereas the antisymmetric contribution is imaginary. It is common to define $\sigma^{\alpha;\beta\gamma}_{\text{shift},C}\equiv \text{Im} \,\sigma^{\alpha;[\beta\gamma]}_{\text{shift}}$ in connection to its relation to circularly polarized light. It is evident that integration over positive frequencies eliminates the dependence on the band gaps $\epsilon_{mn}$ and results in the sum rules
\begin{alignat}{3}
    &\int_0^\infty \!\!\!\!\!d\omega\,\,\sigma^{\alpha;(\beta\gamma)}_\text{shift}(\omega)&&
    =\,-\frac{2\pi e^3}{\hbar^2}\sum_{\underset{n<m}{n,m}}\int_\bk f_{nm}\,\text{Im}\,C^{mn}_{\alpha;(\beta\gamma)} \, ,\\
    &\int_0^\infty \!\!\!\!\!d\omega\,\,\sigma^{\alpha;[\beta\gamma]}_\text{shift}(\omega)&&
    =-\frac{2i\pi e^3}{\hbar^2}\sum_{\underset{n<m}{n,m}}\int_\bk f_{nm}\,\text{Re}\,C^{mn}_{\alpha;[\beta\gamma]} \, ,
\end{alignat}
which are entirely geometric for insulators at sufficiently low temperatures where $f_{nm} \approx 1$ for occupied bands $n$ and unoccupied bands $m$. As pointed out in Ref.~\onlinecite{Wojciech2024}, the sum of the circularly related components of the circular shift current is related to the torsion after frequency integration, that is
\begin{align}
    \int_0^\infty \!\!\!\!\!d\omega\,\,\Big[\sigma^{\alpha;[\beta\gamma]}_\text{shift}(\omega)+\sigma^{\beta;[\gamma\alpha]}_\text{shift}(\omega)+\sigma^{\gamma;[\alpha\beta]}_\text{shift}(\omega)\Big] = -\frac{i\pi e^3}{\hbar^2}\sum_{\underset{n<m}{n,m}}\int_\bk f_{nm}\,\text{Re}\,\Big[T^{mn}_{\alpha;\beta\gamma}+T^{mn}_{\beta;\gamma\alpha}+T^{mn}_{\gamma;\alpha\beta}\Big] \, .
\end{align}

\section{Separation of the third-order geometric tensor $C^{mn}_{\alpha;\beta\gamma}$}

We have a closer look at the symmetrization of $C^{mn}_{\alpha;\beta\gamma}$ in both the band indices and external indices $\beta\leftrightarrow \gamma$ and find
\begin{align}
    C^{(mn)}_{\alpha;(\beta\gamma)} &=\delta_{nm}\frac{1}{2}\big(Q^n_{\beta;\alpha\gamma}+Q^n_{\gamma;\alpha\beta}\big)-\frac{1}{4}\text{tr}\big[\hat e_\beta^{nm}\,\partial_\alpha \hat e_\gamma^{mn}+\hat e_\beta^{mn}\,\partial_\alpha \hat e_\gamma^{nm}+\hat e_\gamma^{nm}\,\partial_\alpha \hat e_\beta^{mn}+\hat e_\gamma^{mn}\,\partial_\alpha \hat e_\beta^{nm}\big] \\
    &=\delta_{nm}\frac{1}{2}\big(Q^n_{\beta;\alpha\gamma}+Q^n_{\gamma;\alpha\beta}\big)-\frac{1}{4}\text{tr}\big[\partial_\alpha\big(\hat e_\beta^{nm}\,\hat e_\gamma^{mn}+\hat e_\beta^{mn}\,\hat e_\gamma^{nm}\big)\big]\\
    &=\delta_{nm}\frac{1}{2}\big(Q^n_{\beta;\alpha\gamma}+Q^n_{\gamma;\alpha\beta}-\partial_\alpha Q^n_{\beta;\gamma}\big)+\frac{1}{2}\partial^{}_\alpha\,Q^{(mn)}_{\beta\gamma} \, , \\
    C^{[mn]}_{\alpha;[\beta\gamma]}&=\frac{1}{2}\,\partial_\alpha\,Q^{[mn]}_{\beta\gamma} \, .
\end{align}
with $Q^n_{\beta;\alpha\gamma} = \text{tr}\big[\hat P_n\partial_\beta \hat P_n\partial_\alpha\partial_\gamma\hat P_n\big]$ and $\hat e^{mn}_\alpha = i\hat P_m(\partial_\alpha \hat P_n)\hat P_n$; more details on their properties and relations are summarized in the companion article \cite{Mitscherling2024}. In particular, the off-diagonal components in the band indices are related to the momentum derivative of the two-state, second-order quantum geometric tensor $Q^{mn}_{\beta\gamma}$. The upper identities imply the following decomposition of the off-diagonal components into
\begin{align}
    C^{mn}_{\alpha;\beta\gamma}=C^{mn}_{\alpha;(\beta\gamma)}+C^{mn}_{\alpha;[\beta\gamma]}=&\frac{1}{2}\partial^{}_\alpha\,Q^{mn}_{\beta\gamma}+C^{(mn)}_{\alpha;[\beta\gamma]}+C^{[mn]}_{\alpha;(\beta\gamma)}\nonumber\\=&\frac{1}{2}\partial^{}_\alpha\,Q^{mn}_{\beta\gamma}+\text{Re}\,C^{mn}_{\alpha;[\beta\gamma]}+i\,\text{Im}\,C^{mn}_{\alpha;(\beta\gamma)} \hspace{10mm} \text{for} \hspace{10mm} m\neq n\, ,
\end{align}
where the physical implications of the last two contributions are the focus of the presented study, see Eq.~\eqref{eqn:Cmn_decomp} in the main text, with a special emphasis on $C^{[mn]}_{\alpha;(\beta\gamma)}$.

\section{Summation over band indices of $Q^{mn}_{\beta\gamma}$ and $C^{mn}_{\alpha;\beta\gamma}$}

It is a natural question whether the geometric quantities $Q^{mn}_{\beta\gamma}$ and $C^{mn}_{\alpha;\beta\gamma}$, defined in Eqs.~\eqref{eqn:Qmn_bc} and \eqref{eqn:Cmn_abc} in the main text, can be expressed entirely in terms of the projector $\hat P_\text{occ} = \sum_{n\in \text{occ}} \hat P_n$ when the indices $m$ and $n$ are traced over the unoccupied and occupied states. The projectors $\hat P_n$ correspond to (degenerate) bands with energies $E_n$. Indeed, we have
\begin{align}
    -\text{tr}\big[\hat P_\text{occ} (\partial_\beta \hat P_\text{occ})(\partial_\gamma \hat P_\text{occ})\big] &= \text{tr}\big[\hat P_\text{occ}\,\big(\partial_\beta(\hat 1-\hat P_\text{occ})\big)\,(\partial_\gamma \hat P_\text{occ})\big] \\[2mm] &= \sum_{\substack{n,n' \in \text{occ} \\ m\in \text{unocc}}} \bigg(\text{tr}\big[\hat P_n \hat P_m (\partial_\beta \hat P_m)(\partial_\gamma \hat P_{n'})\big] + \text{tr}\big[\hat P_n (\partial_\beta \hat P_m)\hat P_m (\partial_\gamma \hat P_{n'})\big]\bigg) \\ &= \sum_{\substack{n,n' \in \text{occ} \\ m\in \text{unocc}}} \delta_{nn'}\,\text{tr}\big[\hat P_n (\partial_\beta \hat P_m)\hat P_m(\partial_\gamma \hat P_{n'})\big] =\sum_{\substack{n\in \text{occ} \\ m\in \text{unocc}}} Q^{mn}_{\beta\gamma} \, ,
\end{align}
where we used $\hat P_n\hat P_m = 0$ for $n\neq m$ and $\hat P_m (\partial_\gamma \hat P_{n'})\hat P_n = (\delta_{mn'}+\delta_{nn'})\hat P_m (\partial_\gamma \hat P_{n'})\hat P_n$. Thus, $Q^{mn}_{\beta\gamma}$ reduces to a ground-state property. In contrast, the first term of $C^{mn}_{\alpha;\beta\gamma}$ yields  
\begin{align}
    \sum_{\substack{n \in \text{occ} \\ m\in \text{unocc}}}\!\!\!\!\! \text{tr}\big[\hat P_n (\partial_\beta \hat P_m)(\partial_\alpha\partial_\gamma \hat P_n)\big]
    =&-\text{tr}\big[\hat P_\text{occ}(\partial_\beta \hat P_\text{occ})(\partial_\alpha\partial_\gamma \hat P_\text{occ})\big] \nonumber \\
    &+\!\!\!\sum_{n\in \text{occ}}\!\!\!\text{tr}\big[\hat P_\text{occ}(\partial_\beta \hat P_\text{occ})(\partial_\alpha \hat P_n)(\partial_\gamma \hat P_n)\big]
    + \!\!\!\sum_{n\in \text{occ}}\!\!\!\text{tr}\big[\hat P_\text{occ}(\partial_\beta \hat P_\text{occ})(\partial_\gamma \hat P_n)(\partial_\alpha \hat P_n)\big] \, ,
\end{align}
where we used the identity for the second-order derivative $\partial_\alpha\partial_\beta \hat P_n = \hat P_n(\partial_\alpha\partial_\beta \hat P_n)+(\partial_\alpha\partial_\beta \hat P_n)\hat P_n + (\partial_\alpha \hat P_n)(\partial_\beta \hat P_n) + (\partial_\beta \hat P_n)(\partial_\alpha \hat P_n)$. We see that the term reduces to the ground state quantity in the first line, since $\text{tr}\big[\hat P_0(\partial_\alpha \hat P_0)(\partial_\beta \hat P_0)(\partial_\gamma \hat P_0)\big] = 0$. Bringing the second term of $C^{mn}_{\alpha;\beta\gamma}$ into the form
\begin{align}
    \sum_{\substack{n \in \text{occ} \\ m\in \text{unocc}}} \text{tr}\big[\hat P_n (\partial_\beta \hat P_m)(\partial_\alpha \hat P_m)(\partial_\gamma \hat P_n)\big] &= \sum_{n\in \text{occ}}\text{tr}\big[\hat P_\text{occ} (\partial_\beta \hat P_\text{occ}) (\partial_\alpha \hat P_n)(\partial_\gamma \hat P_n)\big]\nonumber \\ & +\sum_{m\in \text{unocc}}\text{tr}\big[\hat P_\text{occ}(\partial_\beta \hat P_m)(\partial_\alpha \hat P_m)(\partial_\gamma \hat P_\text{occ})\big] \, ,
\end{align}
using $(\partial_\gamma \hat P_\text{occ})\hat P_\text{occ} = \sum_{n\in \text{occ}}\hat P_n (\partial_\gamma \hat P_n) \hat P_\text{occ} + \sum_{n\in\text{occ}}(\partial_\gamma \hat P_n)\hat P_n$ and $(\partial_\alpha \hat P_n)\hat P_m = - \hat P_n(\partial_\alpha \hat P_m)$ for $n\neq m$, we see that also the existence of multiple unoccupied bands lead to a deviation from the ground state property.

\section{Torsion tensor in projector form}

We note that $C^{(mn)}_{\alpha;[\beta\gamma]}$ has recently been related to the torsion defined as \cite{Ahn2020}
\begin{align}
    C^{mn}_{\gamma;\beta\alpha}-C^{mn}_{\alpha;\beta\gamma}=\text{tr}\big[\hat e^{nm}_\beta\big(\nabla^{}_\alpha\,\hat e^{mn}_\gamma-\nabla^{}_\gamma\,\hat e^{mn}_\alpha-[\hat e^{mn}_\gamma,\hat e^{mn}_\alpha]\big)\big] \equiv T^{mn}_{\beta;\alpha\gamma}
\end{align}
when combined adequately \cite{Wojciech2024}. Using that the commutator vanishes, $[\hat e^{mn}_\gamma,\hat e^{mn}_\alpha]=0$, we see that the cyclic sums are related via
\begin{align}
    C^{(mn)}_{\alpha;[\beta\gamma]}+C^{(mn)}_{\beta;[\gamma\alpha]}+C^{(mn)}_{\gamma;[\alpha\beta]}=\frac{1}{2}\text{Re}\big[T^{mn}_{\alpha;\beta\gamma}+T^{mn}_{\beta;\gamma\alpha}+T^{mn}_{\gamma;\alpha\beta}\big] \, .
\end{align}
The torsion expressed in projectors reads
\begin{align}
    T^{mn}_{\beta;\alpha\gamma} = \text{tr}\Big[\hat P_n (\partial_\beta \hat P_m)\big[(\partial_\alpha\hat P_m)(\partial_\gamma\hat P_n)-(\partial_\gamma \hat P_m)(\partial_\alpha \hat P_n)\big]\Big] \, ,
\end{align}
which we give in Eq.~\eqref{eqn:torsion} in the main text.

\section{Application to transition metal dichalcogenides (TMDs)}

We summarize the basic definitions and techniques, which we use in application to transition metal dichalcogenides presented in the main text. We build on the low-energy models presented in Ref.~\onlinecite{Liu2013}. We consider a monolayer of transition metal dichalcogenides $\text{MX}_2$ placed in the $x$-$y$ plane. The M atoms form a triangular lattice. The presence of the $\text{X}=\text{S},\,\text{Se},\,\text{Te}$ atoms modify the nearest-neighbor (NN) and third-nearest-neighbor (TNN) hopping amplitudes between the $\text{M}$ sites, where we use the NN and TNN models with hopping amplitudes obtained by fitted first-principle band structures in the generalized-gradient approximation (GGA) and local-density approximation (LDA) as presented in Ref.~\onlinecite{Liu2013}. We focus on the spinless case and do not include spin-orbit coupling.

\begin{figure*}[b!]
    \centering
    \includegraphics[width=0.5\linewidth]{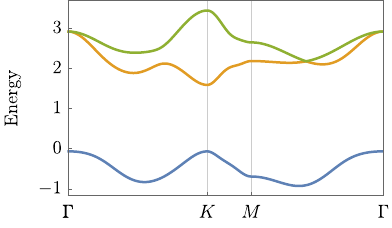} 
    \caption{The band structure on the high-symmetry axes for $\text{MoS}_2$ using the TNN-GGA model. We label the lowest band (blue) as 1 and the two upper bands as 2 (orange) and 3 (green).}
    \label{fig:bandstructureMoS2}
\end{figure*}

\subsection{Basic definitions of the three-band tight-binding models}

We use the lattice constant between neighboring $\text{M}=\text{Mo},\,\text{W}$ as our unit of length setting $a = 1$ but keep it finite in the following definitions for reference. We define the lattice unit vectors as $\ba_1 = a\,(1,0,0)$, $\ba_2 = a\,(1/2, \sqrt{3}/2, 0)$, and $\ba_3 = a_z\,(0,0,1)$ with lattice constant $a_z$ in $z$-direction. The volume of the unit cell is $V = |\ba_3\cdot (\ba_2 \times \ba_1)| = \sqrt{3}\,a^2 a_z/2$. We define the reciprocal lattice vectors 
\begin{align}
    &  \bb_1 = \frac{2\pi}{V} \, \ba_2 \times \ba_3  = \frac{2\pi}{a} \, \bigg(1, -\frac{1}{\sqrt{3}}, 0\bigg) \, , \\[2mm]
    & \bb_2 = \frac{2\pi}{V} \, \ba_3 \times \ba_1 = \frac{2\pi}{a} \,\bigg(0,\frac{2}{\sqrt{3}}, 0\bigg) \, , \\[2mm]
    & \bb_3 = \frac{2\pi}{V} \, \ba_1 \times \ba_2 = \frac{2\pi}{a_z}\, \bigg(0, 0, 1\bigg) \, ,
\end{align}
satisfying $\ba_i\cdot \bb_j = 2\pi\,\delta_{ij}$. The high-symmetry momentum points within this coordinate system are given by 
\begin{align}
    \boldsymbol{\Gamma} = (0,0,0)\, , \hspace{1cm} \text{\bf M} = \frac{1}{2}\,\bb_1 + \frac{1}{2}\,\bb_2 = \frac{2\pi}{a}\bigg(\frac{1}{2},\,\frac{1}{2\sqrt{3}},\,0\bigg) \, , \hspace{1cm} \text{\bf K} = \frac{2}{3}\,\bb_1+\frac{1}{3}\,\bb_2 = \frac{2\pi}{c}\bigg(\frac{2}{3},\,0,\,0\bigg) \, .
\end{align}
We show the band dispersion on the high symmetry axes for $\text{MoS}_2$ using the TNN-GGA model parameters \cite{Liu2013}. We refer to the lowest band (blue) as band 1 and the two upper bands as band 2 (orange) and band 3 (green). It is convenient for numerical evaluation to introduce the coordinate system $\bk(\tilde k_1, \tilde k_2, \tilde k_3) = \tilde k_1\,\bb_1 + \tilde k_2\,\bb_2 + \tilde k_3\,\bb_3$, such that $\tilde k_i \in [0,1)$ covers the Brillouin zone (BZ). The corresponding Jacobian reads $J = 16\pi^3/(\sqrt{3}\, a^2 a_z) = (2\pi)^3/V$, so that the integral over the BZ simplifies to 
\begin{align}
    V\!\!\int_\bk \equiv \frac{V}{(2\pi)^3}\!\!\int_\text{BZ}\!\!\!\!d\bk = \int_0^1 \!\!\!d\tilde k_1 \int_0^1 \!\!\!d\tilde k_2 \int_0^1 \!\!\!d\tilde k_3 \, .
\end{align}
Since the integrand does not depend on $\tilde k_3$, we can set this integral to unity.

\subsection{Variance and skewness of lowest energy band}

We summarize the variance and skewness calculated for the lowest band of TMDs (blue in Fig.~\ref{fig:bandstructureMoS2}). Note that we have $\langle \hat x \, \hat y \rangle_c = 0$ and 
$\langle \hat x^2 \rangle_c = \langle \hat y^2 \rangle_c$ with $\hat x \equiv \hat X_x$ and $\hat y \equiv \hat X_y$ for the variance. For the skewness, we have the relations $\langle \hat x^3 \rangle_c = \langle \hat x \,\hat y^2\rangle_c = 0$ and
$\langle \hat y^3 \rangle_c = - \langle \hat x^2 \, \hat y \rangle_c $. Thus, we focus on the components 
\begin{alignat}{3}
    &V\!\!\int_\bk\text{Im}\,Q^{11}_{yy}&&=V\!\!\int_\bk\text{Re}\Big[\text{tr}\big[\hat P_1 (\partial_y \hat P_1)(\partial_y \hat P_1)\big]\Big] \, ,\\
    &V\!\!\int_\bk\text{Im}\,C^{11}_{y;yy}&&=V\!\!\int_\bk\text{Im}\Big[\text{tr}\big[\hat P_1(\partial_y \hat P_1)(\partial_y \partial_y \hat P_1)\big]\Big] \, .
\end{alignat}
We numerically evaluate the projectors and their derivatives as described in the companion paper \cite{Mitscherling2024}. We summarize the results in Tab.~\ref{tab:TableVarianceSkewness} for six TMDs using the four different model parameters as given in Ref.~\onlinecite{Liu2013}.

\setlength\extrarowheight{1mm}
\begin{table}[b!]
    \centering
    \setlength{\arrayrulewidth}{0.2mm}
    \begin{tabular}{ c | c c c c c c}
        $V\!\!\int_\bk\text{Im}\,Q^{11}_{yy} $ & \hspace{3mm} $\text{MoS}_2$ \hspace{3mm} & \hspace{3mm} $\text{WS}_2$\hspace{3mm} & \hspace{3mm} $\text{MoSe}_2$ \hspace{3mm} &\hspace{3mm}  $\text{WSe}_2$\hspace{3mm}  &\hspace{3mm} $\text{MoTe}_2$ \hspace{3mm}  &\hspace{3mm}  $\text{WTe}_2$ \hspace{3mm} \\[1mm]\hline \hline
        NN (GGA)    & $0.172$       & $0.175$       & $0.169$       & $0.170$       & $0.165$       & $0.165$    \\[1mm]\hline
        NN (LDA)    & $0.171$       & $0.174$       & $0.168$       & $0.169$       & $0.164$       & $0.164$    \\[1mm]\hline \hline
        TNN (GGA)   & $0.278$       & $0.281$       & $0.295$       & $0.305$       & $0.321$       & $0.343$    \\[1mm]\hline
        TNN (LDA)   & $0.278$       & $0.287$       & $0.307$       & $0.320$       & $0.272$       & $0.357$    \\[1mm]\hline \hline
    \end{tabular}
    \\[5mm]
    \begin{tabular}{ c | c c c c c c}
        $V\!\!\int_\bk\text{Im}\,C^{11}_{yyy} $ & \hspace{3mm} $\text{MoS}_2$\hspace{3mm}  & \hspace{3mm} $\text{WS}_2$\hspace{3mm} & \hspace{3mm} $\text{MoSe}_2$ \hspace{3mm} & \hspace{3mm} $\text{WSe}_2$ \hspace{3mm} & \hspace{3mm} $\text{MoTe}_2$ \hspace{3mm}  & \hspace{3mm} $\text{WTe}_2$ \hspace{3mm} \\[1mm]\hline \hline
        NN (GGA)  & $0.016$       & $0.028$       & $0.011$       & $0.023$       & $0.003$       & $0.015$    \\[1mm]\hline
        NN (LDA)    & $0.019$       & $0.030$       & $0.013$       & $0.025$       & $0.005$       & $0.018$    \\[1mm]\hline \hline
        TNN (GGA)   & $-0.119$       & $-0.123$       & $-0.130$       & $-0.139$       & $0.022$       & $-0.157$    \\[1mm]\hline
        TNN (LDA)   & $-0.121$       & $-0.129$       & $-0.147$       & $-0.158$       & $-0.119$       & $-0.176$    \\[1mm]\hline \hline
    \end{tabular}
    \caption{ The variance (upper table) and skewness (lower table) for the six TMDs calculated using the four parameters corresponding to NN and TNN obtained via GGA and LDA density functional theory data \cite{Liu2013}. 
    }
    \label{tab:TableVarianceSkewness}
\end{table}

\subsection{Angle dependence}

We analyze the angle dependence of the variance and skewness by evaluating the partial derivatives for varying direction 
\begin{align}
    \hat X_\phi = \cos(2\pi\phi) \,\hat x + \sin(2\pi\phi) \, \hat y \, ,
\end{align}
which corresponds to an adjusted momentum unit vector $\mathbf{e}_\phi = \big(\!\cos(2\pi\phi),\sin(2\pi\phi)\big)$ in the numerical evaluation. We give explicitly calculated variance and skewness for $\text{MoS}_2$ using the TNN-GGA model parameter \cite{Liu2013} as black dots in Fig.~\ref{fig:angledependence}. We identify that the variance is angle independent, whereas the skewness $\langle (\hat X_\phi)^3\rangle \propto \sin(6\pi \phi)$ follows the three-fold rotational symmetry $C_3$ of the underlying lattice structure.

\begin{figure*}[b!]
    \centering
    \includegraphics[width=0.42\linewidth]{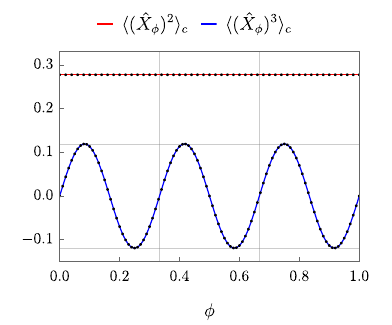}  
    \caption{The variance (red) and skewness (blue) for $\text{MoS}_2$ using the TNN GGA model as a function of directional angle for $\hat X_\phi = \cos(2\pi\phi)\,\hat x + \sin(2\pi\phi)\,\hat y$. We compare the explicitly calculated results using Eqs.~\eqref{eqn:variance} and \eqref{eqn:skewness} in the main text (black dots) and the identified angle dependence $\langle (\hat X_\phi)^2\rangle \propto 1$ and $\langle (\hat X_\phi)^3\rangle \propto \sin (6\pi \phi)$ with maximal value at $y$-direction.}
    \label{fig:angledependence}
\end{figure*}

\begin{table}[b!]
    \begin{tabular}{ c | c c c c c c}
        $V\!\!\int_\bk\text{Im}\,C^{21}_{y;yy}$ & \hspace{3mm} $\text{MoS}_2$\hspace{3mm}  & \hspace{3mm} $\text{WS}_2$\hspace{3mm} & \hspace{3mm} $\text{MoSe}_2$ \hspace{3mm} & \hspace{3mm} $\text{WSe}_2$ \hspace{3mm} & \hspace{3mm} $\text{MoTe}_2$ \hspace{3mm}  & \hspace{3mm} $\text{WTe}_2$ \hspace{3mm} \\[1mm]\hline \hline
        NN (GGA)  & $0.012$       & $-0.001$       & $0.017$       & $0.006$       & $0.018$       & $0.006$    \\[1mm]\hline
        NN (LDA)    & $0.011$       & $-0.004$       & $0.018$       & $0.005$       & $0.018$       & $0.005$    \\[1mm]\hline \hline
        TNN (GGA)   & $0.075$       & $0.078$       & $0.080$       & $0.087$       & $0.008$       & $0.094$    \\[1mm]\hline
        TNN (LDA)   & $0.075$       & $0.080$       & $0.089$       & $0.096$       & $0.092$       & $0.100$    \\[1mm]\hline \hline
    \end{tabular}
    \\[5mm]
    \begin{tabular}{ c | c c c c c c}
        $V\!\!\int_\bk\text{Im}\,C^{31}_{y;yy}$ & \hspace{3mm} $\text{MoS}_2$\hspace{3mm}  & \hspace{3mm} $\text{WS}_2$\hspace{3mm} & \hspace{3mm} $\text{MoSe}_2$ \hspace{3mm} & \hspace{3mm} $\text{WSe}_2$ \hspace{3mm} & \hspace{3mm} $\text{MoTe}_2$ \hspace{3mm}  & \hspace{3mm} $\text{WTe}_2$ \hspace{3mm} \\[1mm]\hline \hline
        NN (GGA)  & $-0.012$       & $-0.011$       & $-0.012$       & $-0.012$       & $-0.009$       & $-0.012$    \\[1mm]\hline
        NN (LDA)    & $-0.012$       & $-0.011$       & $-0.012$       & $-0.012$       & $-0.009$       & $-0.013$    \\[1mm]\hline \hline
        TNN (GGA)   & $0.053$       & $0.062$       & $0.065$       & $0.072$       & $-0.002$       & $0.072$    \\[1mm]\hline
        TNN (LDA)   & $0.060$       & $0.069$       & $0.077$       & $0.082$       & $0.043$       & $0.082$    \\[1mm]\hline \hline
    \end{tabular}
    \\[5mm]
    \begin{tabular}{ c | c c c c c c}
        $V\!\!\int_\bk\text{Im}\,\Delta C^{11}_{y;yy}$ & \hspace{3mm} $\text{MoS}_2$\hspace{3mm}  & \hspace{3mm} $\text{WS}_2$\hspace{3mm} & \hspace{3mm} $\text{MoSe}_2$ \hspace{3mm} & \hspace{3mm} $\text{WSe}_2$ \hspace{3mm} & \hspace{3mm} $\text{MoTe}_2$ \hspace{3mm}  & \hspace{3mm} $\text{WTe}_2$ \hspace{3mm} \\[1mm]\hline \hline
        NN (GGA)  & $0.017$       & $0.017$       & $0.016$       & $0.017$       & $0.013$       & $0.009$    \\[1mm]\hline
        NN (LDA)    & $0.018$       & $0.016$       & $0.019$       & $0.018$       & $0.014$       & $0.010$    \\[1mm]\hline \hline
        TNN (GGA)   & $0.009$       & $0.017$       & $0.015$       & $0.021$       & $0.028$       & $0.009$    \\[1mm]\hline
        TNN (LDA)   & $0.014$       & $0.020$       & $0.019$       & $0.020$       & $0.017$       & $0.006$    \\[1mm]\hline \hline
    \end{tabular}
    \caption{The integrated geometric contributions $C^{21}_{y;yy}$ (upper table) and $C^{31}_{y;yy}$ (middle table) for the six TMDs for the four model parameters \cite{Liu2013}, which contribute the linear shift current sum rule. The deviation from skewness is summarized in the table at the bottom.
    }
    \label{tab:C21C31Summary}
\end{table}

\subsection{Individual contributions to the sum rule}

We calculate the contributions to the linear shift current sum rule assuming a filled lowest band. We calculate the relevant geometric quantities for the $yyy$-component via
\begin{alignat}{3}
    &V\!\!\int_\bk\text{Im}\,C^{21}_{y;yy}&&=V\!\!\int_\bk\text{Im}\Big[\text{tr}\big[\hat P_1(\partial_y \hat P_2)(\partial_y\partial_y \hat P_1)\big]+\text{tr}\big[\hat P_1(\partial_y \hat P_2)(\partial_y \hat P_2)(\partial_y \hat P_1)\big]\Big] \, ,\\
    &V\!\!\int_\bk\text{Im}\,C^{31}_{y;yy}&&=V\!\!\int_\bk\text{Im}\Big[\text{tr}\big[\hat P_1(\partial_y \hat P_3)(\partial_y\partial_y \hat P_1)\big]+\text{tr}\big[\hat P_1(\partial_y \hat P_3)(\partial_y \hat P_3)(\partial_y \hat P_1)\big]\Big] \, ,
\end{alignat}
using the numerical procedure described in the companion article \cite{Mitscherling2024}. Similarly, the interband correction to the skewness as given in Eq.~\eqref{eqn:sum_Cmn} is calculated via
\begin{alignat}{3}
    &V\!\!\int_\bk\text{Im}\,\Delta C^{11}_{y;yy}&&=V\!\!\int_\bk\text{Im}\Big[\text{tr}\big[\hat P_1\big((\partial_y \hat P_2)(\partial_y \hat P_2)+(\partial_y \hat P_3)(\partial_y \hat P_3)\big)(\partial_y \hat P_1)\big]\Big] \, .
\end{alignat}
We summarize the individual results for six TMDs in Tab.~\ref{tab:C21C31Summary} using the four different model parameters presented in Ref.~\onlinecite{Liu2013}. 

\subsection{Sum rule results}

We combine the individual contribution to the sum rule via
\begin{align}
    &\text{Sum rule} \equiv -V\!\!\int_\bk\text{Im}\,C^{21}_{y;yy} - V\!\!\int_\bk\text{Im}\,C^{31}_{y;yy} \, ,
\end{align}
which is related to the values presented in Fig.~\ref{fig:TMDresults} via $2\pi/V \times (\text{Sum rule})$. We calculate the ratio  
\begin{align}
    &R \equiv \frac{V\!\!\int_\bk\text{Im}\,C^{11}_{y;yy}}{\text{Sum rule}}
\end{align}
to identify the relative contribution of the skewness to the sum rule. We summarize the results in Tab.~\ref{tab:sumRule}, where we include only the TNN model parameters. Note the significantly different results for $\text{MoTe}_2$ using the TNN-GGA parameters (in brackets).

\begin{figure*}[t!]
    \centering
    \includegraphics[width=0.3\linewidth]{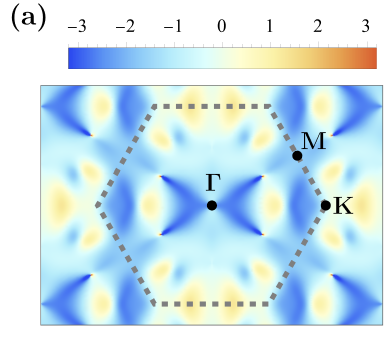}  
    \includegraphics[width=0.3\linewidth]{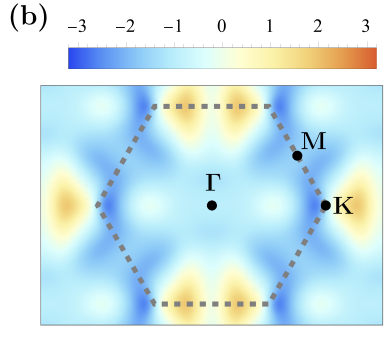}  
    \includegraphics[width=0.3\linewidth]{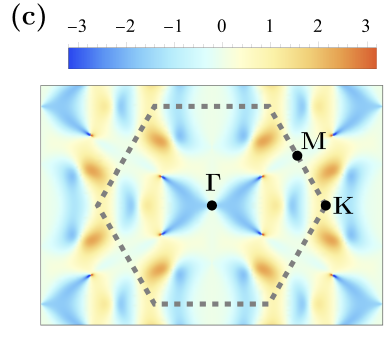} 
    \caption{The momentum-resolved contributions calculated for $\text{MoS}_2$ using the TNN-GGA model parameters. (a) The integrant of the shift current sum rule in units of $e^3/\hbar^2$ given by $-2\pi/V\,\text{Im}\big[C^{21}_{y;yy}(\bk)+C^{31}_{y;yy}(\bk)\big]$. (b) The skewness of the lowest band contributing to the shift current sum rule given by $2\pi/V\,\text{Im}\big[C^{11}_{y;yy}\big]$. (c) The multi-band geometry contributing to the linear shift current sum rule given by $-2\pi/V\,\text{Im}\big[C^{21}_{y;yy}(\bk)+C^{31}_{y;yy}(\bk)+C^{11}_{y;yy}(\bk)\big]$.  }
    \label{fig:MomentumResolvedSumRule}
\end{figure*}

\subsection{Momentum-resolved evaluation of the individual contributions}

We have a closer look at the momentum dependency of the different contributions. We summarize the momentum-resolved sum rule, the skewness, and the deviation between these for $\text{MoS}_2$ evaluated for TNN-GGA model parameters in Fig.~\ref{fig:MomentumResolvedSumRule}. 

\subsection{Torsion}

The torsion calculated via Eq.~\eqref{eqn:torsion} has nonvanishing components $T^{mn}_{x;xy} = -T^{mn}_{x;yx} \neq 0$ and $T^{mn}_{y;xy} = -T^{mn}_{y;yx} \neq 0$ as shown by explicit calculation for various model parameters. The corresponding sum rule for the circular shift current vanishes due to the restriction to the 2d plane since 
\begin{align}
    &T^{mn}_{x;xy}+T^{mn}_{x;yx}+T^{mn}_{y;xx}=T^{mn}_{x;xy}-T^{mn}_{x;yx} = 0 \, ,\\
    &T^{mn}_{y;xy}+T^{mn}_{x;yy}+T^{mn}_{y;yx}=T^{mn}_{y;xy}-T^{mn}_{y;xy} = 0 \, .
\end{align}

\begin{table}
    \begin{tabular}{ c c | c c c c c c}
        &  & \hspace{3mm} $\text{MoS}_2$\hspace{3mm}  & \hspace{3mm} $\text{WS}_2$\hspace{3mm} & \hspace{3mm} $\text{MoSe}_2$ \hspace{3mm} & \hspace{3mm} $\text{WSe}_2$ \hspace{3mm} & \hspace{3mm} $\text{MoTe}_2$ \hspace{3mm}  & \hspace{3mm} $\text{WTe}_2$ \hspace{3mm} \\[1mm]\hline \hline
        $\text{Sum rule} $ & TNN (GGA)                  & $-0.128$       & $-0.140$       & $-0.145$       & $-0.159$       & $(-0.006)$       & $-0.166$    \\[1mm]\hline
         & TNN (LDA)                                    & $-0.135$       & $-0.149$       & $-0.166$       & $-0.178$       & $-0.135$       & $-0.182$    \\[1mm]\hline \hline
        $R\,[\%] $ & TNN (GGA)   & $93.0$       & $87.9$       & $89.7$       & $87.4$       & $(366.7)$       & $94.6$    \\[1mm]\hline
         & TNN (LDA)                                    & $89.6$       & $86.6$       & $88.6$       & $88.8$       & $88.1$       & $96.7$    \\[1mm]\hline \hline
    \end{tabular}
    \caption{ The sum rule result and the relative contribution from the skewness for the six TMDs for the TNN models parameters \cite{Liu2013}.
    }
    \label{tab:sumRule}
\end{table}

\section{Minimal three-band model for skewness and its interband correction: \\the Stub Rice-Mele model}
We study the Rice-Mele model \cite{Rice1982} combined with the one-dimensional stub lattice \cite{Neves2024} given by the Hamiltonian  
\begin{align}
    \hat H = \sum_j \Big[&\big(1+\delta\big)\big(\hat c^\dagger_{j,A}\hat c^{}_{j,B}+\hat c^\dagger_{j,B}\hat c^{}_{j,A}\big)+\big(1-\delta\big)\big(\hat c^\dagger_{j+1,A}\hat c^{}_{j,B}+\hat c^\dagger_{j,B}\hat c^{}_{j+1,A}\big)\nonumber\\&+m\,\big(\hat c^\dagger_{j,A}\hat c^{}_{j,A}-\hat c^\dagger_{j,B}\hat c^{}_{j,B}\big) + \Delta\,\big(\hat c^\dagger_{j,A}\hat c^{}_{j,C}+\hat c^\dagger_{j,C}\hat c^{}_{j,A}\big)\Big] \, ,
\end{align}
where $\delta\in [-1,1]$ captures the asymmetry in the intra- and intercell hopping, $m$ captures the relative energy between sites A and B, and $\Delta$ captures the coupling to the stubs attached to the A sites. For $\Delta=0$, the model reduces to the Rice-Mele model. For $t=m=0$ and $\Delta=1$, the model correspond to the one-dimensional Stub lattice. We set the sites A and C at the origin and the site B at the center of the unit cell of size 1, i.e., $r_A=r_C=0$ and $r_B=\frac{1}{2}$. The Fourier transformation to lattice momentum,
\begin{align}
    \hat c^\dagger_{j,\alpha} = \int\frac{dk}{\sqrt{2\pi}}\,\hat c^\dagger_{k,\alpha}\,e^{-ik(R_j+r_\alpha)} \, ,
\end{align}
leads to the Bloch Hamiltonian matrix
\begin{align}
    \hat H(k) = \begin{pmatrix} m & 2\cos\frac{k}{2}+2i\delta\sin \frac{k}{2} & \Delta \\ 2\cos \frac{k}{2}-2i\delta\sin \frac{k}{2} & - m & 0 \\ \Delta &  0 & 0 \end{pmatrix}
\end{align}
in the basis $\hat c^\dagger_k \equiv \big(\hat c^\dagger_{k,A},\,\hat c^\dagger_{k,B},\hat c^\dagger_{k,C}\big)$. We sort the three bands as $E_1(k)<E_2(k)<E_3(k)$. Using the trigonometric solutions for real roots of the depressed cubic equation, we obtain analytic formulas for the band energies. The bands are separated as long as either $\delta$, $m$, or $\Delta$ is nonzero. We focus on the geometry of the lowest band in analogy to the application to TMDs. We use the formalism presented by Graf and Pi\'echon \cite{Graf2021} built upon the decomposition into SU(3) Gell-Mann matrices to derive analytic expressions for the band projectors $\hat P_n(k)$, which we apply to the three geometric invariants of interest. In particular, we calculate the variance and skewness via 
\begin{alignat}{3}
    &\text{Var}(\delta,m,\Delta)&&=\int_{-\pi}^\pi\frac{dk}{2\pi}\,\,\text{Re}\Big[\text{tr}\big[\hat P_1 (\partial_k \hat P_1)(\partial_k \hat P_1)\big]\Big] \, ,\\
    &\text{Skew}(\delta,m,\Delta)&&=\int_{-\pi}^\pi\frac{dk}{2\pi}\,\,\text{Im}\Big[\text{tr}\big[\hat P_1(\partial_k \hat P_1)(\partial_k \partial_k \hat P_1)\big]\Big] \, .
\end{alignat}
and the interband correction via
\begin{alignat}{3}
    &\text{Corr}(\delta,m,\Delta)&&=\int_{-\pi}^\pi\frac{dk}{2\pi}\,\,\text{Im}\Big[\text{tr}\big[\hat P_1\big((\partial_k \hat P_2)(\partial_k \hat P_2)+(\partial_k \hat P_3)(\partial_k \hat P_3)\big)(\partial_k \hat P_1)\big]\Big] \, .
\end{alignat}
\subsection{Skewness of the Rice-Mele model}
The model reduces to the Rice-Mele model \cite{Rice1982} for $\Delta=0$, which decouples the highest and lowest band from the trivial flat band at zero energy. Thus, the  model reduces to a two-band model plus a trivial atomic chain. We obtain
\begin{align}
    \text{Var}(\delta,m,0) &= \frac{1+\delta^2}{8\sqrt{(4+m^2)(m^2+4\delta^2)}} \, , \\ 
    \text{Skew}(\delta,m,0) &= \frac{m \,\delta}{4(4+m^2)\pi \sqrt{m^2+4\delta^2}}E\Big[\frac{4(\delta^2-1)}{m^2+4\delta^2}\Big] \, ,\\[2mm]
    \text{Corr}(\delta,m,0) &= 0 \, ,
\end{align}
with the complete elliptic integral of the second kind $E(x) = \int_0^{\pi/2} \sqrt{1-x \sin^2\theta}\,d\theta$. As expected for a two-band model, the interband correction vanishes. For $\delta=m=0$, the model reduces to a trivial one-dimensional chain with closed gap at $k=\pi$ arising from the doubled unit cell. The variance $\text{Var}(\delta,m,0)$ is finite for either nonzero $t$ and $m$, whereas the skewness $\text{Skew}(\delta,m,0)$ requires both parameters to be nonzero. We can interpret this as follows: The inequality in the intra- and intercell hopping $\delta\neq 0$ leads to a dimer formation on the A-B bonds, which is skewed by the energy difference between them due to $m\neq 0$. This is particularly evident in the extreme dimerized limit without intercell hopping for $\delta=1$, where the third standardized moment reads 
\begin{align}
    \frac{\text{Skew}(1,m,0)}{\text{Var}(1,m,0)^{3/2}} = m \, .
\end{align}
Note that third standardized moment changes sign for the other extreme dimerized limit $\delta=-1$ as expected.
\subsection{Skewness of the stub-SSH lattice}
The model reduces to the one-dimensional stub lattice \cite{Neves2024} for $m=0$ and $\delta=0$. We keep $\delta\neq 1$ in the following, leading to a combination of the Su-Schrieffer-Heeger (SSH) model and the stub model. We obtain 
\begin{align}
    \text{Var}(\delta,0,\Delta)&= \frac{4\Delta^2+\big(1+\delta^2\big)\big(\Delta^4+4\delta^2(2+\Delta^2)\big)}{4\big((4+\Delta^2)(4\delta^2+\Delta^2)\big)^{3/2}} \, ,\\
    \text{Skew}(\delta,0,\Delta)&= -\frac{\delta\Delta^2\big(-12-12\delta^6+\Delta^2+5\Delta^4+\Delta^6+\delta^4(28+\Delta^2)+\delta^2(28+30\Delta^2+5\Delta^4)\big)}{4\big((4+\Delta^2)(4\delta^2+\Delta^2)\big)^{5/2}} \, ,\\
    \text{Corr}(\delta,0,\Delta)&=-\frac{\delta\Delta^2\big(12+12\delta^6+7\Delta^2+\Delta^4+\delta^4(4+7\Delta^2)+\delta^2(4+2\Delta^2+\Delta^4)\big)}{2\big((4+\Delta^2)(4\delta^2+\Delta^2)\big)^{5/2}}
\end{align}
The variance $\text{Var}(\delta,0,\Delta)$ is finite for either nonzero $\delta$ or $\Delta$, whereas the skewness $\text{Skew}(\delta,0,\Delta)$ and the interband correction $\text{Corr}(\delta,0,\Delta)$ requires both parameters to be nonzero. As for the Rice-Mele model, the formation of dimers is essential. In contrast, the skewness arises via the third site C. This suggets that a nonzero interband contribution $\text{Corr}(\delta,0,\Delta)$ is an indicator of a three-site (or three-band) origin of the skewness in contrast to a two-site origin of the Rice-Mele model. The formulas simplify in the extreme dimerized limit with $\delta=1$, 
\begin{alignat}{3}
    \text{Var}(1,0,\Delta) &= \frac{2+\Delta^2}{2(4+\Delta^2)^2} &&\approx \frac{1}{16}\big(1-\frac{1}{16}\Delta^4\big) \, ,\\
    \text{Skew}(1,0,\Delta) &=-\frac{\Delta^2(2+\Delta^2)}{4(4+\Delta^2)^3} &&\approx -\frac{\Delta^2}{128}\big(1-\frac{1}{4}\Delta^2\big) \,, \\
    \text{Corr}(1,0,\Delta) &=-\frac{\Delta^2}{(4+\Delta^2)^3}&&\approx -\frac{\Delta^2}{64}\big(1-\frac{3}{4}\Delta^2\big) \, .
\end{alignat}
We see that the finite size of the dimer $(1/4)^2$ arising from the finite size $r_B-r_A=1/2$ is barely affected by the hybridization with site C. We find that the interband correction dominates with respect to the skewness for sufficiently small $\Delta$. In this regime, we expect a strong deviation from the sum rule given in the main text. We capture the quality of the sum rule via 
\begin{align}
    \label{eqn:Rfunction}
    R(\delta,m,\Delta) = \frac{\text{Skew}(\delta,m,\Delta)}{\text{Skew}(\delta,m,\Delta)-\text{Corr}(\delta,m,\Delta)} \approx 1+\frac{\text{Skew}(\delta,m,\Delta)}{\text{Corr}(\delta,m,\Delta)} \, .
\end{align}
Indeed, we have $R(1,0,\Delta)\approx-1$ for small $\Delta$, such that the interband correction reverses the sign of the frequency-integrated linear shift current, strongly deviating from the sum rule given in the main text. When the coupling between sites A and C far exceeds the coupling between sites A and B, we find
\begin{align}
    \frac{\text{Skew}(1,0,\Delta)}{\text{Var}(1,0,\Delta)^{3/2}} &= -\frac{\Delta^2}{\sqrt{2}\sqrt{2+\Delta^2}} \underset{|\Delta|\rightarrow\infty}{\rightarrow} -\frac{1}{\sqrt{2}}|\Delta| \, , \\
    \frac{\text{Corr}(1,0,\Delta)}{\text{Var}(1,0,\Delta)^{3/2}} &= -\frac{2\sqrt{2}\Delta^2}{(2+\Delta^2)^{3/2}}\underset{|\Delta|\rightarrow\infty}{\rightarrow} -\frac{2\sqrt{2}}{|\Delta|} \, .
\end{align}
Since the model effectively reduces to a two-band model involving the sites A and C, the vanishing interband correction term is expected. 
\subsection{Interplay of the two- and three-site mechanism for skewness}
We study the competing aspect of the two distinct mechanisms for skewness. For the fully dimerized limit $\delta=1$, we obtain 
\begin{align}
    \text{Var}(1,m,\Delta) &\approx \frac{1}{4(4+m^2)}+\frac{m\Delta^2}{2(4+m^2)^{\frac{5}{2}}} \, ,\\
    \text{Skew}(1,m,\Delta) &\approx \frac{m}{8(4+m^2)^\frac{3}{2}}+\frac{(m^2-2)\Delta^2}{4(4+m^2)^3} \, ,\\
    \text{Corr}(1,m,\Delta) &\approx -\frac{\Delta^2}{(4+m^2)^3} \, .
\end{align}
for sufficiently small $\Delta$. We clearly see the interplay between the onsite energy difference $m$ of sites A and B and the hybridization between sites A and C, confirming the interpretation of the underlying mechanism beyond the aforementioned special limits. To shed further light onto the competing mechanism, we calculate the standardized moment finding
\begin{align}
    \frac{\text{Skew}(1,m,\Delta)}{\text{Var}(1,m,\Delta)^{3/2}} &\approx m-\frac{\Delta^2}{\sqrt{4+m^2}}\, , \\
    \frac{\text{Corr}(1,m,\Delta)}{\text{Var}(1,m,\Delta)^{3/2}} &\approx \frac{8\Delta^2}{(4+m^2)^\frac{3}{2}} \, .
\end{align}
Lowering the energy on site B with respect to site A $(m>0)$ leads to a positive skewness, whereas the hybridization between A and C implies a negative skewness. Thus, we find a strongly suppressed skewness for $m\approx \frac{1}{2}\Delta^2$ in coexistence with of a nonzero interband correction. We conclude that the interband correction can dominate the frequency-integrated linear shift current. In Fig.~\ref{fig:S4}, we present the interplay beyond the small $\Delta$ and the perfect dimer limit $(|\delta|<1)$. We find that for equal onsite energies both the skewness and interband correction are of equal importance but may yield different signs dependent on the level of dimerization; see Fig.~\ref{fig:S4}~(a,b). When the onsite energy difference is the dominating mechanism, the skewness far exceeds the interband correction until the stub hybridization and onsite energy are of same order; see Fig.~\ref{fig:S4}~(c,d). The quality of the sum rule quantified by Eq.~\eqref{eqn:Rfunction} for the Stub Kane-Mele model is assured for a large parameter regime as long as the two-site mechanism is the dominant origin of skewness; see Fig.~\ref{fig:S5}.
\begin{figure*}[t!]
    \centering
    \includegraphics[width=0.38\linewidth]{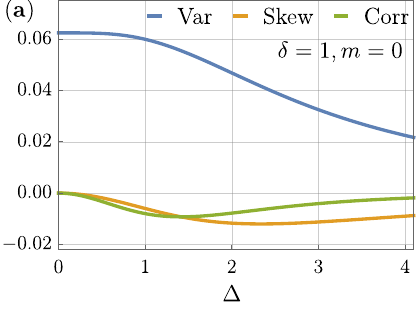} \hspace{3mm}
    \includegraphics[width=0.38\linewidth]{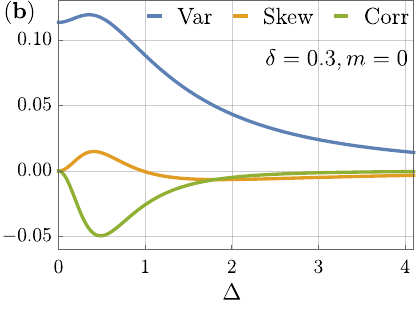}
    \includegraphics[width=0.38\linewidth]{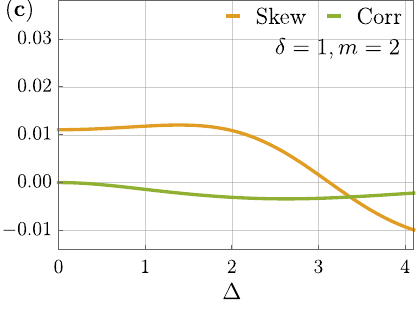} \hspace{3mm}
    \includegraphics[width=0.38\linewidth]{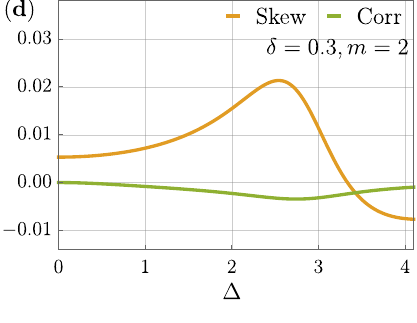}
    \caption{The variance, skewness, and interband correction of the Stub Kane-Mele model as a function of the stub hybridization $\Delta$ in the perfectly dimerized limit (a,c) and dispersive limit (b,d). The interplay between $\Delta$ and $m$ (c,d) can lead to sign change in the skewness.
    }
    \label{fig:S4}
\end{figure*}
\begin{figure*}[t!]
    \centering
    \includegraphics[width=0.38\linewidth]{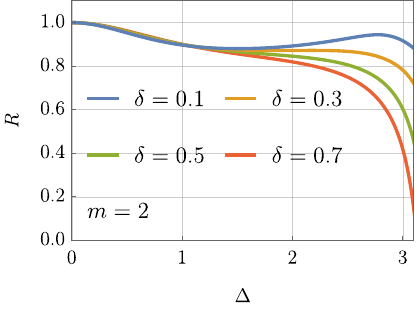}
    \caption{The quality of the sum rule as a function of $\Delta$ for different levels of dimerization. We see that a good approximation of the sum rule by the skewness $(R\approx 90\%)$ can be achieved within the Stub Kane-Mele model for a broad parameter range of $\delta$ and $\Delta$, when the two-site energy difference $(m=2)$ is the dominant mechanism for skewness. 
    }
    \label{fig:S5}
\end{figure*}

\newpage

\end{widetext}

\end{document}